\documentclass[aps,twocolumn,floatfix]{revtex4}

\usepackage{mciteplus}

\usepackage{times}
\usepackage{graphics}% Include figure files
\usepackage{epsfig,calc}% Include figure files
\usepackage{bm}  %boldmath

\usepackage{amsmath,amssymb,chemarr}

\def\btt1{{\tt$\backslash$\string1}}%

\def\AmS{{\protect\the\textfont2
        A\kern-.1667em\lower.5ex\hbox{M}\kern-.125emS}}

\citeindextrue

\newcommand{\wsub}{W_\text{sub}}
\newcommand{\kt}{k_\text{B}T}
\newcommand{\rhat}{\hat{r}}

\newcommand{\fh}{f_\text{f}}
\newcommand{\fnh}{f_\text{nf}}
\newcommand{\uh}{W_\text{int}}

\newcommand{\ub}{U_\text{bias}}
\newcommand{\zh}{Z_\text{f}}
\newcommand{\znh}{Z_\text{nf}}
\newcommand{\xic}{\xi_\text{c}}
\newcommand{\rc}{R_\text{c}}
\newcommand{\Lg}{L_\text{g}}

\begin{document}

\title{The pair potential of colloidal stars}

\author{F. Huang,$^1$ K. Addas,$^{2,}$}

\altaffiliation[Present address: ]{Department of Physics, The American University in Cairo, P.O. Box 74 New Cairo 11835, Egypt}

\author{ A. Ward,$^1$\\
N. T. Flynn,$^3$ E. Velasco,$^4$ M. F. Hagan,$^1$ Z. Dogic,$^1$ and
S. Fraden$^1$}

\affiliation{
$^1$Department of Physics, Brandeis University, Waltham, MA 02454, USA\\
$^2$Rowland Institute at Harvard, Cambridge, MA 02142, USA\\
$^3$Department of Chemistry, Wellesley College, Wellesley, MA 02481, USA\\
$^4$Departamento de F\'{\i}sica Te\'{o}rica de la Materia Condensada, Universidad Aut\'{o}noma de Madrid, E-28049 Madrid, Spain
}

% and Instituto de Ciencia de Materiales Nicol\'{a}s Cabrera,

\date{\today}

\begin{abstract}
We report on the construction of colloidal stars: 1 $\mu$m polystyrene
beads grafted with a dense brush of 1 $\mu$m long and 10 nm wide semi-flexible
filamentous viruses. The pair interaction potentials of colloidal stars are measured using an experimental implementation of umbrella sampling, a technique originally developed in computer simulations in order to probe rare events. The influence of ionic
strength and grafting density on the interaction is measured. Good agreements are found between the measured interactions and theoretical predictions based upon the osmotic pressure of counterions.
\end{abstract}

\maketitle

Polyelectrolyte brushes of flexible polymers have been the subject of many theoretical~\cite{Pincus91,Russel89,Ballauff06} and experimental~\cite{Ballauff06} studies. Recently focus has shifted to semiflexible brushes~\cite{Kegler07} for which the persistence length $P$ is large compared to the monomer separation, but small compared to their contour length $L$, or $P << L$. In contrast, here we investigate brushes with $P\sim L$. The grafted brushes consist of bacteriophage M13 viruses, which are rodlike, semiflexible charged polymers of length $L=880$ nm, diameter $D=6.6$ nm, and persistence length $\sim 2 \mu$m~\cite{Khalil07,*Song91}. The bare, linear charge density of M13 is high; $\sim 7 e^- / \mbox{nm}$.

\begin{figure}
\epsfig{file=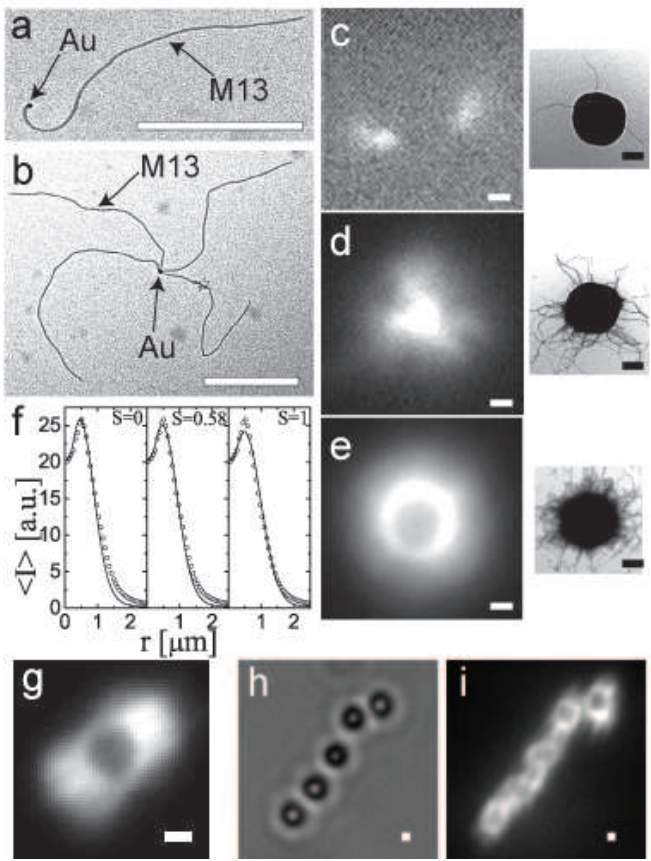,width=8.6cm}
\caption{
    (a) and (b): TEM images of 10 nm Au-bound M13 viruses of different nanoarchitectures.
    (c)-(e): TEM (right panel) and fluorescence (left panel) images of labeled phage grafted to unlabeled 1 $\mu$m PS beads with
    varying grafting densities. (c) 3 phages/bead. (d) 38 phages/bead. (e) 135 phages/bead.
    (f): Radially-averaged fluorescent intensity profiles of the phage-grafted
    bead. Symbols: experiment; Solid
    curve: theoretical calculation with varying orientational order parameters ($S$) of anchored rods. (g): Fluorescent image of colloidal star in a M13 nematic (in contrast to (e) where the solvent is isotropic). The ``hair" grafted to the bead is ``combed" parallel to the director by the nematic. (h): Brightfield image of colloidal stars associating end-to-end in a M13 nematic. (i): Fluorescent image of (h). The combed stars associate in chains aligned parallel to the nematic director with surfaces separated by a micron. Bare spheres in a nematic also assemble into chains, but with surfaces in contact. The scale bars are 500 nm.}
\label{M13-Au-HB}
\end{figure}

In this letter, we describe ``colloidal stars'', which are analogous to star polymers~\cite{Pincus91,Jusufi02,*Dominguez-Espinosa08,Jusufi04,Ballauff06}, constructed by grafting genetically
engineered M13 viruses~\cite{Mao04} to polystyrene spheres. These stiff brushes represent a new class of stars. The M13 are rigid enough to form liquid crystals~\cite{Dogic06}, but when grafted to a sphere remain flexible enough to be distorted by the director field, as shown in Fig.~\ref{M13-Au-HB}. These colloidal
structures are characterized by fluorescent microscopy,
transmission electron microscopy (TEM) and fluorometry.
The interaction potential is probed using laser tweezers. To extract the steeply varying pair-potential we develop a new experimental protocol based on the computer simulation method known as umbrella sampling~\cite{Torrie77}, but modified to increase the protocol's efficiency under experimental constraints. This new method allows the measurement of potentials much greater in magnitude than done perviously with line traps~\cite{Crocker99,Lin01}. We find that the measured potential of the colloidal stars can be modeled as arising from the osmotic pressure of the counter-ions, which is in several fold excess of the repulsion due to rod excluded volume.

M13 bacteriophage was grown and purified as described
elsewhere~\cite{Maniatis89}. The M13 capsid protein pIII, present only on one end of the virus, was modified in order to display cysteine residues. We achieved this through making use of the Ph.D.-C7C Phage Display Peptide Library
(M13-C7C, New England Biolabs, Beverly, MA)~\cite{Mao04}. We were able to create colloidal stars with core sizes varying from 10 nm to 1 $\mu$m.
Fig.~\ref{M13-Au-HB}a and b show M13-C7C viruses conjugated with 10-nm colloidal Au particles (Ted Pella, Redding, CA).
In this article, we focus on the colloidal star constructed by attaching the engineered phages to a 1 micron diameter polystyrene sphere. This was done using the following procedure:
First, 230 $\mu$l of 8.8 mg/ml M13-C7C was reduced with 2 $\mu$l of 0.18 mg/ml TCEP
(Tris(2-carboxyethyl)phosphine) for 15 min. This M13-C7C solution
was mixed with 2 $\mu$l of 19 mM maleimide-PEO$_2$-biotin (Pierce, Rockford, IL) for 1 h in 20 mM
phosphate buffer at $p$H = 7.0. The phage solution was dialyzed
extensively against phosphate buffer to remove excess biotin and
the $p$H was readjusted to 8.0. Subsequently, the phages were mixed
for 1 h with 1 mg/ml Alexa Fluor$^\circledR$ 488 carboxylic acid succinimidyl
ester (Molecular Probes, Eugene, OR), and centrifuged four times at $170,000g$ for 1 h to
remove free dye molecules. 0.5 mg/ml of the fluorescently-labelled viruses were then
 incubated with $0.5\%$(w/v) straptavidin-coated polystyrene beads of diameter $d=0.97\pm 0.02$
$\mu$m (Bangs Laboratories, Fishers, IN) for 24 hours at room
temperature. To the suspension 0.05 mg/ml $\alpha$-casein (Sigma, St. Louis, MO)
 was added and the whole mixture was centrifuged twice at $20,000g$ for 10 min. Finally,
 the pellet was resuspended in 100 $\mu$l of phosphate
buffer (5 mM, $p$H 8.0) and stored at $4 ^{\circ}$C.
%Fluorometry was performed on the suspension to evaluate
The number of the
sphere-bound viruses was determined using a fluorescence spectrophotometer
(F-2000, Hitachi, Tokyo, Japan). By varying the stoichiometric
ratio of biotinylated viruses to straptavidin-coated beads
we created star polymers of different grafting
densities as revealed by both fluorescence and TEM images (Fig.~\ref{M13-Au-HB}(c-e)).
 Fluorescent images were taken on a fluorescence microscope (TE2000-U, Nikon) equipped with a $100\times$ oil-immersion objective and a
cooled CCD camera (CoolSnap HQ, Roper Scientific).
The TEM samples, stained with $2\%$ uranyl acetate, were imaged with a 268 microscope (Morgagni, FEI Company, Hillsboro, OR), operating at 80 kV.

At the grafting density of 135 phages/bead (Fig.~\ref{M13-Au-HB}e), the
anchored dye-labelled rods form a spherically symmetric corona around the bead
with a radially-averaged intensity (RAI) profile shown in
Fig.~\ref{M13-Au-HB}(f). We model the phage-grafted bead as hard rods
anchored to the sphere with a Gaussian angle distribution, which is
centered around the surface normal. The
diffraction-limited fluorescent image of the colloidal star was computed by convolving the
distribution of the rod's segments with the theoretical 3D point
spread function (PSF) of the microscope~\cite{Born80}. As can be seen from Fig.~\ref{M13-Au-HB}(f) the calculated RAI profiles are insensitive to
the orientational order parameter of the anchored rods $S=\frac{1}{2}\langle3\cos^2\theta-1\rangle$, where $\theta$ is the angle between the rod and the surface normal. However, the best fits were for intermediate order parameters.

%The interaction energy of two colloidal particles
%$U(r)$ (equivalent to their Helmholtz free energy)
The free energy as a function of separation between two colloidal particles $\uh(r)$ (the potential of mean force) can be
determined up to an additive offset by the Boltzmann relation,
$P(r)\sim\exp[-\uh(r)/\kt]$. Experimentally this is accomplished by measuring the
probability $P(r)$ of finding the particles at a separation $r$. However, for states of even
moderate repulsive interaction energies $P(r)$ becomes very small. As a result,
infrequent visitation of improbable states leads to poor statistics and
errors in the determination of $P(r)$ which limited the magnitude of measured potentials in previous implementations of line traps, or single bias potentials to about ~6 $\kt$~\cite{Crocker99,Lin01}. In this paper the maximum measured potential is 40~$\kt$, but we estimate that potentials several times this value are feasible with the laser power and optical resolution of our instrument.

\begin{figure}
\epsfig{file=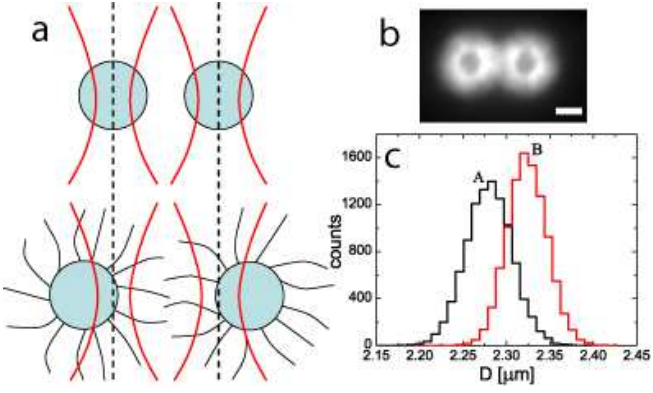,width=8.6cm}
\caption{
    (color online). Excluded volume interaction of anchored rods.
    (a) the schematic  and (b) the fluorescence image of phage-grafted beads in optical
    traps. (c) separation
    histograms of (A) bare beads and (B) phage-grafted beads for the same trap locations. The scale bar in (b) is 1 $\mu$m.
    }
\label{HBschem&hist}
\end{figure}

% MFH added some stuff here
We achieve these measurements by employing the method of umbrella sampling, in which a biasing force is used to enhance sampling of rare configurations; results are then re-weighted to obtain the physical probability distribution~\cite{Torrie77}.  Specifically, we place two colloidal stars (Fig.~\ref{HBschem&hist}b) in separate laser traps and measure the histogram of separation distances between the colloids.  The measurement is performed in a series of windows, each of which uses a different separation distance between the minima of the two laser traps. In each window the stars fluctuate about the minimum of a total potential resulting from a combination of the dual traps and interparticle star potential. Only 6 $\kt$ of each of the total potentials is sampled and each minimum has a different energy, but here we show how the total potentials from overlapping windows can be combined to produce a single interparticle pair-potential of large range and magnitude. For the protocol typically used in simulations, results from different windows would be simultaneously re-weighted and stitched together to obtain a continuous function for the probability $P(r)$ using the weighted histogram analysis method (WHAM)~\cite{Ferrenberg89,*Kumar92,*Roux95}.  However, the biasing potential is a function of two coordinates because the position of each bead is controlled by a separate trap.  The number of independent measurements required for a particular level of statistical accuracy using WHAM rises exponentially with the number of dimensions of the biasing potential (even if the probability is projected onto a single coordinate).  We overcome this limitation as follows.

Our goal is to measure the interaction potential, $\uh(r)$ with $r\equiv x_2-x_1$, between a pair of functionalized particles sitting at positions $(x_1,x_2)$ in a bias potential (laser trap) of strength $\ub(x_1,x_2)$. We achieve this goal by performing two experiments (Fig. 2a).
In one experiment we place two colloidal stars in two separate laser traps and in the other experiment we place two bare colloids in the same two traps. For both experiments we measure the separation histogram of the colloids as a function of the trap separation.
The potential of mean force, $\wsub$, is then obtained by subtracting the results from each experiment.
\begin{equation}
\wsub(\rhat)/\kt = -\log[\fh(\rhat)]+\log[\fnh(\rhat)]
\label{eq:wsub}
\end{equation}
with $\fh(\rhat)$ and $\fnh(\rhat)$ the fraction of measured displacements that fall within the histogram bin associated with the displacement
value $\rhat$ for functionalized and  non-functionalized beads, respectively.  While this subtraction method has been used in previous experiments~\cite{Crocker99,Lin01}, we rigorously prove its validity here and show how to implement it over multiple windows.

The fractions of measured displacements are governed by the Boltzmann distribution and given by
\begin{equation}
\fnh(\rhat)=\znh^{-1}\int dx_1 \int dx_2 e^{-\ub(x_1,x_2)/\kt} \delta(x_1-x_2-\rhat)
\label{eq:fnh}
\end{equation}
and
\begin{eqnarray}
\fh(\rhat)&=&\zh^{-1}\int dx_1 \int dx_2 e^{-\ub(x_1,x_2)/\kt}\nonumber \\
&&\times e^{-\uh(x_2-x_1)/\kt}\delta(x_1-x_2-\rhat)
\label{eq:fh}
\end{eqnarray}
with $\delta(r)$ the Dirac delta function and the normalization factors are given by
\begin{eqnarray}
\znh&=&\int dx_1 \int dx_2 e^{-\ub(x_1,x_2)/\kt}\nonumber \\
\zh&=&\int dx_1 \int dx_2 e^{-\ub(x_1,x_2)/\kt}e^{-\uh(x_2-x_1)/\kt}
\label{eq:zs}
\end{eqnarray}

We change the integration variables to $x_1$ and $r\equiv x_2-x_1$ and integrate over $r$ to obtain
\begin{eqnarray}
\fnh(\rhat)&=& \znh^{-1}\int dx_1 e^{-\ub(x_1,\rhat)/\kt} \nonumber \\
\fh(\rhat)&=& \zh^{-1}e^{-\uh(\rhat)/\kt}\int dx_1 e^{-\ub(x_1,\rhat)/\kt}
\label{eq:intr}.
\end{eqnarray}
Inserting this result into Eq.~\ref{eq:wsub} gives the calculated potential of mean force:
\begin{equation}
\wsub(\rhat) = \uh(\rhat) + \kt\log(\zh/\znh)
\label{eq:error}.
\end{equation}

We see that $\wsub(\rhat) = \uh(\rhat)$ plus a constant. As discussed above, the strength of the laser traps, $\ub(x_1,x_2)$, is such that the colloids sample only a small range and therefore only a small piece of the interaction potential $\uh(\rhat)$ is obtained. To determine a wider range of $\uh(\rhat)$ the laser trap separation is varied and $\wsub(\rhat)$ is obtained anew. Although the constant term is different for each separation of the traps, the entire potential can be stitched together to within a single additive constant by assuming that $\uh(\rhat)$ is continuous.

\begin{figure*}
\epsfig{file=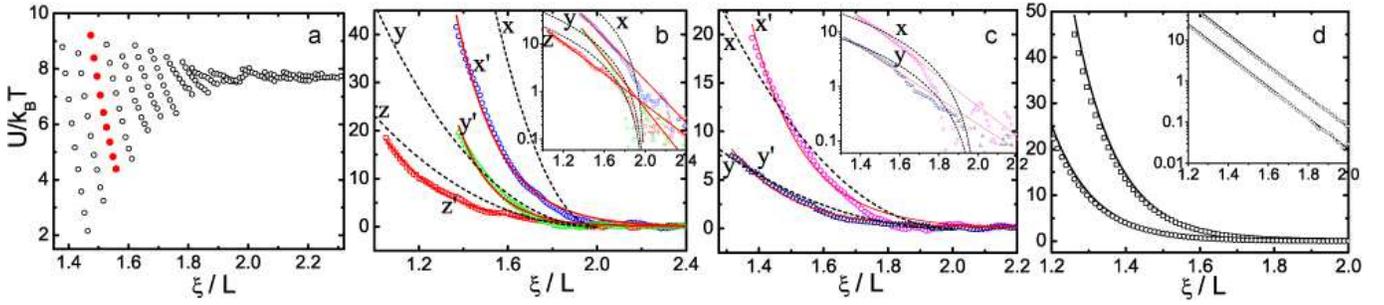, width=18cm}
\caption{
    (color online). (a) A set of interaction potentials of M13-grafted micron-sized polystyrene spheres acquired from each umbrella window, with the ionic strength $I=14$ mM and the grafting density $\sigma=135$ phages/bead. The solid symbols indicate the potential extracted from the histograms shown in Fig.~\ref{HBschem&hist}c.
    (b) Pair interaction potentials of colloidal stars
    at varying solution ionic strengths with $\sigma=135$ phages/bead. Symbols:
    experiment; dashed lines: theory; solid lines: single exponential
    fits. ($\circ$,x') and (x): 2.8 mM; ($\vartriangle$,y') and (y): 14 mM; ($\square$,z') and (z): 28 mM.
    (c) Pair potentials at different grafting densities with $I=14$ mM. ($\circ$,x') and (x): 135 phages/bead;
    ($\vartriangle$,y') and (y): 80 phages/bead.  (d) Interaction potentials $U/\kt=Be^{-10(r-2.2)}$ employed in the Langevin dynamics simulation (Solid lines) and bias potential $\ub(\mathbf{r}_1,\mathbf{r}_2)=\frac{1}{2}k_1|\mathbf{r}_{1}-\mathbf{r}_{c1}|^2+\frac{1}{2}k_2|\mathbf{r}_{2}-\mathbf{r}_{c2}|^2$.
     Pair potentials extracted using the umbrella sampling (empty symbols).
     $B= 6$ (circle) and 20 (square). Insets: data replotted to facilitate comparison.
     $\xi$ is the separation between the surfaces of spheres and $L$ the virus length.}
\label{exp&theo-potential}
\end{figure*}

As a check of this implementation of the umbrella sampling algorithm, we used computer simulations to model the experiment. The results validating this method are shown in
Fig.~\ref{exp&theo-potential}d.

The experimental system is shown schematically in Fig.~\ref{HBschem&hist}a. The fluorescence image of trapped beads is shown in Fig.~\ref{HBschem&hist}b.
Optical tweezer setup is built around the inverted fluorescence microscope. A single laser beam is time-shared between two points
via a pair of orthogonally oriented paratellurite (\mbox{TeO$_2$})
acousto-optic deflectors (AOD, Intra-Action, Bellwood, IL). About 30 mW of a 1064-nm laser (Laser Quantum, Cheshire, UK) is projected onto the back focal plane of
an oil-immersion objective ($100\times$, N.A.=1.3, Nikon)
and subsequently focused into the sample chamber.  Spheres are trapped 5 $\mu$m away from the surface to minimize possible wall effects.  We choose a
set of umbrella window potentials by systematically varying the
locations of the traps' centers $\mathbf{r}_{c1}$ and
$\mathbf{r}_{c2}$. For each window potential, six minutes of video
are recorded for a pair of phage-grafted beads, and the separation
probability distribution, $\fh(\rhat)$ is obtained. It is a simple Gaussian if the separation is large and the beads are not interacting. The distances between trap positions are selected so that there are sufficient overlaps between
adjacent positions. We collected data for $\sim30$ different trap positions with 50 nm increments in separation to cover a wide range of the interparticle potential. Under identical conditions (microscope
illumination, laser power, sample buffer, etc.), the experiment
was repeated immediately for a pair of streptavidin-coated PS
beads without attached virus to measure $\fnh(\rhat)$. For all experiments, statistically independent
configurations of beads were sampled at 30 frames/sec. We analyzed the
video images using a custom program written in the language IDL~\cite{crocker-JCIS96}. By constructing a histogram of
center-center separations on $10^{4}$ images in each
window, we found clear differences between the separation
probability distributions of virus-grafted beads
$\fh(\rhat)$ and bare beads $\fnh(\rhat)$
(Fig.~\ref{HBschem&hist}c).

%\section{Results}

Fig.~\ref{exp&theo-potential}b shows the interaction potentials
measured between two M13-grafted microspheres with varying
solution ionic strengths. The interactions are all purely
repulsive.
The potential decays to zero as the distance between sphere
surfaces increases beyond twice the virus length. There is a
strong dependence of the pair potential on the ionic strength of
the surrounding medium.  A decrease in the solution ionic strength
leads to increased interaction between spheres grafted with charged rods.
We compare the interaction potential between
microspheres at grafting densities of 80 and 135 viruses per sphere
(Fig.~\ref{exp&theo-potential}c). The increase in density by $68\%$
increases the pair-potential by a factor of 2.6, but does not change its functional form.

We calculated the interaction potential arising from the osmotic pressure of counterions trapped within the grafted layers based on the mean field calculation theory of Jusufi~\cite{Jusufi02,*Dominguez-Espinosa08,Jusufi04}, except modified for the case where the density of fixed charges on the grafted rods is small compared to the salt concentration. In particular, the densities of positive and negative ions within the grafted layer $\rho{\pm}(r)$  are given by $\rho{\pm}\approx \rho_\text{s} \pm 0.5 \rho_{
\text{f}} (r)$, with $\rho_\text{s}$  the salt concentration  and $\rho_{\text{f}} (r)=\lambda_\text{f} N_\text{f}/(4 \pi r^2)$  the fixed concentration of negative charges  on the grafted rods, with $N_\text{f}$ the number of rods per colloid and $\lambda_\text{f}=1.7 e^- / \text{nm}$ an adjustable parameter for the linear charge density renormalized by condensation. The counterion excess free energy is calculated by integrating over the volume of the grafted layer $\int_{0}^{\pi}d\theta \sin \theta\int_{\rc}^{\rc+\Lg(\theta,\xic)} dr r^2  \rho_\text{f}(r)^2/(2 \rho_\text{s})$ with $\rc = 0.5$ $\mu$m the core radius, $\theta$ the angle with the center to center vector for the pair of colloids, and $\xic=\xi+2\rc$ the center to center distance. Following the interaction geometry depicted in Fig. 5 of Ref.~\cite{Jusufi04}, the height of the grafted layer is $\Lg=L$ for $\theta\ge\theta_0$ and $\Lg=\xic/(2 \cos\theta)-\rc$ for $\theta<\theta_0$ with $\cos\theta_0=0.5\xic/(L+\rc)$. We have assumed $\rho_\text{f}\ll \rho_\text{s}$ and that the height of the unperturbed grafted layer is equal to the length of a virus, $L=880$ nm (i.e. we neglect rod orientational fluctuations). Theory and experiment are compared in Figs.~\ref{exp&theo-potential}b \& c; the agreement is rather good considering five measured potentials are fit with one value of the effective charge.

We also calculated the interaction due to the excluded volume of grafted rods based on the Onsager second virial expansion of the free energy~\cite{Onsager49}. For two rods, each with a specified orientation, we find the pairwise excluded area, or the space of relative grafting locations for which the rods overlap. The calculated interaction potential due to rod excluded volume was significantly smaller compared to the interaction due to counterion osmotic pressure; the latter interaction agrees well with the measured interaction potentials.

%\section{discussion}
In conclusion, the umbrella sampling method is applied to extract
the pair potential of the colloidal stars which are trapped with
optical tweezers. The method allows for measurement of potentials of the order of 100~$\kt$, an energy much greater than previously measured with line traps. The large measured repulsive energy between colloidal stars is consistent with the osmotic pressure of counter-ions between the charged brush, while a second virial theory based on the Onsager approximation significantly
underestimates the pair potential.  The construction of colloidal star polymers from
genetically engineered viruses opens the possibility of a
systematic study of hybrid colloidal materials exhibiting complex phase behaviors.

Acknowledgement. We thank Dr. Kirstin Purdy for the initial development of the star colloid, and Dr. Chen Xu for TEM support. Financial support of this work came from NSF (DMR-0444172), NSF (DMR-0705855) and NSF-MRSEC (DMR-0820492).

%\bibliographystyle{apsrevM}
%\bibliography{HB_bib}

\ifx\mcitethebibliography\mciteundefinedmacro
\PackageError{apsrevM.bst}{mciteplus.sty has not been loaded}
{This bibstyle requires the use of the mciteplus package.}\fi

\end{document}